\documentclass[manuscript]{aastex}

\def\arcsa#1#2{$#1^{\prime\prime}_{^\textrm{.}}#2$}
\def\arcs#1{$#1^{\prime\prime}$}

\shorttitle{Shaping of Multipolar PPNe by Bullets}
\shortauthors{Huang, Lee, Moraghan, \& Smith}

\begin{document}

\title{The Shaping of the Multipolar Pre-Planetary Nebula CRL 618 by
Multi-directional Bullets}

\author{Po-Sheng Huang\altaffilmark{1,2}, Chin-Fei Lee\altaffilmark{1,2}, Anthony Moraghan\altaffilmark{1}, and Michael Smith\altaffilmark{3}}

\altaffiltext{1}{Academia Sinica Institute of Astronomy and Astrophysics, P.O. Box 23-141, Taipei 106, Taiwan; posheng@asiaa.sinica.edu.tw}
\altaffiltext{2}{Institute of Astrophysics, National Taiwan University, Taipei 106, Taiwan}
\altaffiltext{3}{Centre for Astrophysics and Space Science, University of Kent, Canterbury CT2 7NH, UK}


\begin{abstract}

In order to understand the formation of the multipolar structures of the
pre-planetary nebula (PPN) CRL 618, we perform 3D simulations using a
multi-directional bullet model.  The optical lobes of CRL 618 and fast
molecular outflows at the tips of the lobes have been found to have similar
expansion ages of $\sim$ 100 yr.  Additional fast molecular outflows were
found near the source along the outflow axes with ages of $\sim$ 45 yr,
suggesting a second episode of bullet ejections.  Thus, in our simulations,
two episodes of bullet ejections are assumed.  The shaping process is
simulated using the ZEUS-3D hydrodynamics code that includes molecular and
atomic cooling.  In addition, molecular chemistry is also included to
calculate the CO intensity maps.  Our results show the following: (1)
Multi-epoch bullets interacting with the toroidal dense core can produce the
collimated multiple lobes as seen in CRL 618.  The total mass of the bullets
is $\sim$ 0.034 M$_{\sun}$, consistent with the observed high-velocity CO
emission in fast molecular outflows.  (2) The simulated CO $J=3$--2
intensity maps show that the low-velocity cavity wall and the high-velocity
outflows along the lobes are reasonably consistent with the observations. 
The position-velocity diagram of the outflows along the outflow axes shows a 
linear increase of velocity with distance, similar to the observations.  The 
ejections of these bullets could be due to magneto-rotational explosions or 
nova-like explosions around a binary companion.

\end{abstract}


\section{INTRODUCTION}

Pre-planetary nebulae (PPNe) are transient objects between the asymptotic
giant branch (AGB) phase and the planetary nebula (PN) phase in the
evolution of low- to intermediate-mass stars.  In less than a thousand
years, PPNe will transform into PNe when the central stars evolve to hot
white dwarfs and photoionize their envelopes.  Most PPNe are aspherical and
found to possess bipolar or multipolar lobes \citep{CS95,S01,S07} with 
kinematic ages less than 1000 yr \citep{B01}.  Bipolar lobes can be
produced by fast winds ejected from around central stars near the end of the
AGB phase.  However, multipolar PPNe may not be produced in the same way,
because the fast winds, even if precessing, have difficulty producing the
multipolar morphology.

CRL 618 is one of the most well-studied multipolar PPNe, and thus a good
candidate for our study of the mass-loss process in multipolar PPNe.  In the
$Hubble$ $Space$ $Telescope$ ($HST$) images, multipolar optical lobes were
observed in the east-west direction \citep{TG02}.  These optical lobes are
expanding rapidly away from the center \citep{S02}.  Infrared observations
in H$_2$ \citep{C03}, and (sub)millimeter observations in CO $J=2$--1
\citep{S04}, $J=3$--2 \citep{L13a,L13b}, and $J=6$--5 \citep{N07} revealed
fast molecular outflows along the optical lobes.  A dense core was detected
around the central star \citep{L13a} surrounded by a tenuous spherical halo
\citep{S04}.

Two-dimensional hydrodynamical simulations have attempted to reproduce the
lobe morphology of CRL 618.  \citet{LS03} and \citet{L09} used a collimated
fast wind (CFW) model to produce one of the well-defined lobes in the west
by using the ZEUS-2D hydrodynamics code.  In the CFW model, the PPN lobes
are a result from the interaction of the fast winds with the surrounding AGB
halo.  The CFW can reproduce the structures and kinematics of the outflow
lobe.  However, the CFW may need to be more confined like a cylindrical jet
in order to produce enough shock emission.

On the other hand, \citet{D08} showed that a massive clump (bullet) model
can also produce the structure and kinematics of a lobe in CRL 618. 
Recently, \citet{B13} measured proper motions of the optical lobes and found
that the fingertips of the lobes have the same expansion ages of $\sim$
100$\pm$15 yr.  Therefore, these optical lobes likely all resulted from a
spray of bullets in different directions.  As a result, \citet{B13} followed
up with the bullet model and included the periodical density enhancement in
the AGB halo in order to produce the observed ringlike structures in the
lobe.

Later, the fast molecular outflows at the tips of the optical lobes are also
found to have similar expansion ages to the optical lobes \citep{L13b}. 
This suggests that the optical lobes and the fast molecular outflows likely
all resulted from the same spray of bullets.
 Additional fast molecular outflows are found near the source
along the outflow axes with ages of $\sim$ 45$\pm$25 yr, suggesting a second
episode of bullet ejections.  But the origin of the bullets is still
unknown.  The bullet ejections could be due to magneto-rotational explosions
\citep{M06} or nova-like explosions \citep{L13a}.

More recently, \citet{R14} adopted a precessing jet model \citep{V12,V13} to
reproduce the multipolar morphology of CRL 618.  If the jet source is also
pulsing and the precession period is an integer multiple of the pulsing
period, the jet can produce multiple lobes.  But this model can only produce
point-symmetric lobes.  In order to reproduce the asymmetric lobes as seen
in CRL 618, a complicated ``asymmetrical jet ejection mechanism" \citep{M12}
is required \citep{V14}.

In this paper, in order to understand the formation of the multipolar
structures of CRL 618, we perform 3D simulations using a multi-directional
bullet model.  Unlike previous bullet models, our AGB halo has
parameters close to those derived from the observations of CRL 618.  In
addition, a dense core around the center is also included, as in the
observations.  The inclusion of the dense core is important to derive the
masses of the bullets accurately, because the resulting velocity of the
bullets depends on the interactions of the bullets with the dense core.  For
the first attempt, we only model the lobes in the east of CRL 618.  As 
suggested in \citet{L13b}, two episodes of bullet ejections are assumed, 
first with two bullets to produce the two major lobes, and second with three 
bullets to produce the fast molecular outflows near the source.  The
numerical settings and assumptions are given in section 2; we compare the 3D
simulations and simulated CO intensity maps with observations in section 3;
in section 4 we discuss the results and possible origin of bullets; and we
summarize this work in section 5.

\section{NUMERICAL SETTINGS} \label{NumSet}

We perform the simulations of the bullet model using the ZEUS-3D code, which
is a grid-based computational fluid dynamics code \citep{C96,C10}.  The code
has been expanded to include molecular and atomic cooling by \citet{S97}. 
In addition, molecular chemistry is also included to calculate the fraction
of molecular hydrogen and atomic hydrogen \citep{SR03}.  The material is
assumed to consist of mostly molecular hydrogen.  Helium is also included
with the number density $n_{\textrm{\scriptsize He}}=0.1
n_{\textrm{\scriptsize H}}$, where $n_{\textrm{\scriptsize H}}$ is the number
density of the hydrogen nuclei.  The CO-to-H$_2$ relative abundance is
assumed to be 2$\times$10$^{-4}$, as adopted in \citet{S04}.

In our simulations, a Cartesian coordinate system is used.  The $x$-axis is
the polar axis.  The $y$- and $z$-axes are in the equatorial plane,
perpendicular to the polar axis.  We use 576$\times$400$\times$400 cells
with a resolution of 12.5 AU per cell to cover from 0 to 7200 AU in the
$x$-axis, $-$2500 to 2500 AU in the $y$-axis, and $-$2500 to 2500 AU in the
$z$-axis.  We adopt a distance $D=900$ pc for CRL 618 \citep{SS04}.  The
resolution of 12.5 AU corresponds to an angular resolution $\sim$
\arcsa{0}{014}, high enough to show the details of the lobes and the
outflows to be compared with the $HST$ \citep[e.g., up to \arcsa{0}{07}
in][]{B13} and SMA observations \citep[e.g., up to \arcsa{0}{5} in][]{L13a}
of CRL 618.

In our simulations, a number of bullets are ejected from the central source,
interacting with the surrounding to produce the multipolar lobes.  The
surrounding consists of an AGB halo with a dense core at the center.

\subsection{Bullets}

\begin{figure}
\centering{
\includegraphics[scale=0.4]{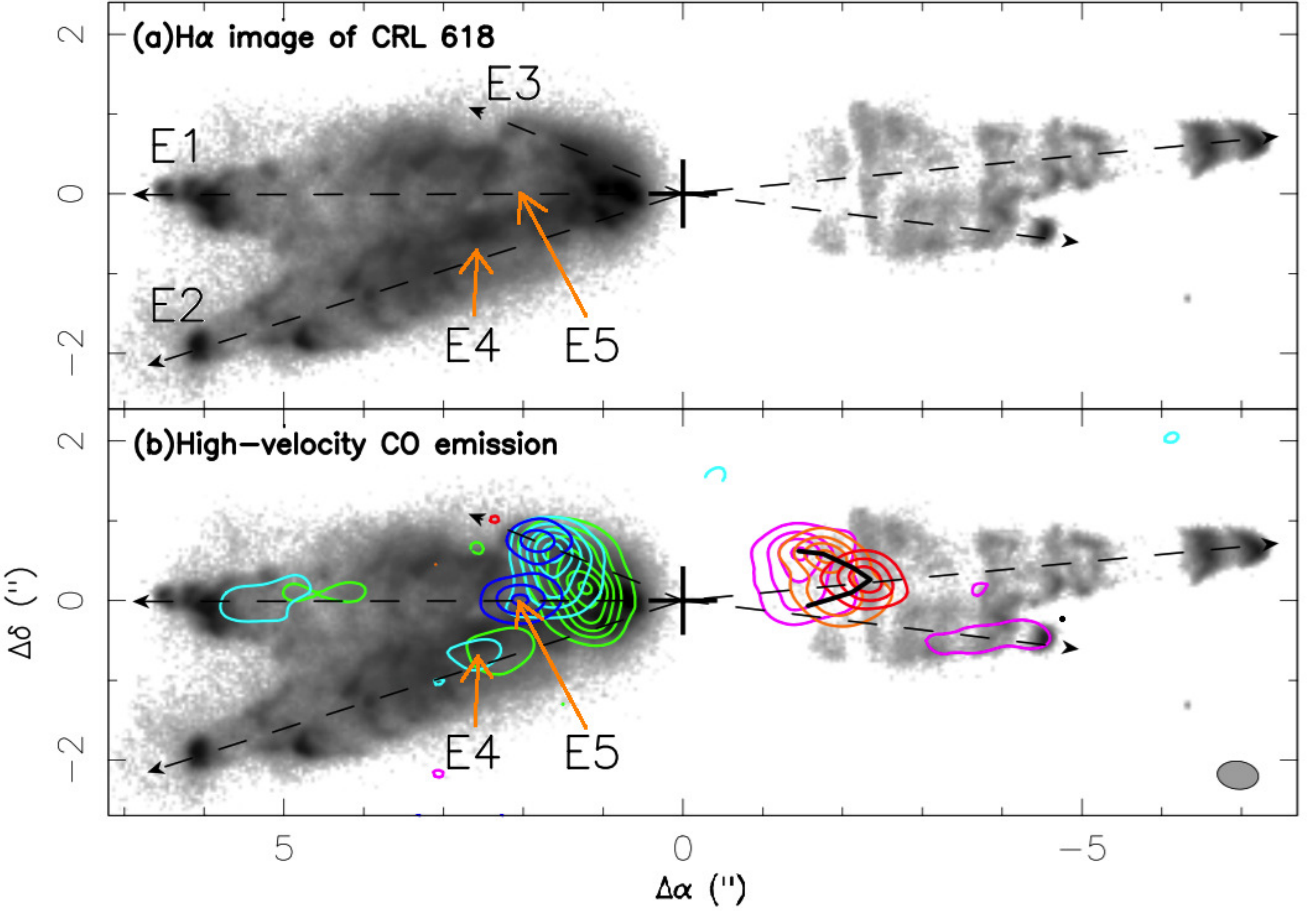}
}
\caption{Upper panel shows the H$\alpha$ image of CRL 618, taken by the 
$HST$ in 2009 August (Camera$=$WFC3/UVIS1, Filter$=$F656N, exposure time$=$560 
s, pixel size$=$\arcsa{0}{04}).  The eastern lobes are labeled with E1 to E5.  
Lobes E1, E2 and E3 are clearly detected in the H$\alpha$ image.  Lobes E4 and E5 can not be identified in the H$\alpha$ image but they have been observed in the high-velocity CO emission \citep{L13b}, shown in the lower panel.   
\label{F1}}
\end{figure}

Figure \ref{F1}a shows the optical lobes of CRL 618 in the H$\alpha$ image, 
taken by the $HST$ in 2009 August.  The outflow lobes labeled with E1, E2 and E3 
are clearly detected in optical.  The labels E4 and E5 indicate the positions of 
the two molecular outflow lobes detected in the high-velocity CO emission map 
\citep{L13b}, as shown in Figure \ref{F1}b.  
In our models, bullets are used to produce the optical lobes and the fast 
molecular outflows along the lobes.  According to the observations \citep{L13b}, 
there would be two episodes of bullet ejections.  In Model 1, two bullets 
(bullets 1 and 2) are ejected from the central source in the first episode at a 
simulation time of $\sim$ 90 yr ago in order to produce the two major lobes, E1 
and E2, in the east of CRL 618.  In order to avoid the collision between the two 
bullets, the second bullet is ejected 6 yr after the first bullet.  In Model 2, 
we add a second episode of three bullets (bullets 3, 4, and 5) at $\sim$ 40 yr 
after the first episode, in order to produce the additional fast outflows (lobes 
E3, E4, and E5) near the source in the east of CRL 618.  Again, in order to 
avoid the collision between the bullets, we assume that bullets 3, 4, and 5 are 
ejected one by one with a time delay of 6 yr.  Notice that lobe E3 was assumed 
to have a similar dynamical age to lobes E1 and E2 in \citet{B13}.  Here it is 
assumed to have a similar dynamical age to lobes E4 and E5, as in \cite{L13b}.  
This assumption is reasonable because lobe E3 has a similar length and radial 
velocity to lobe E5.

All the bullets are assumed to be ejected from the center. They are
cylindrical with a diameter $d_{\textrm{\scriptsize b}}=200$ AU and a length
$l_{\textrm{\scriptsize b}}=200$ AU.  In the observations of CRL 618 by
\citet{S04}, most of the gas in the fast outflows are at temperatures of
$\sim$ 200 K.  Therefore, we set the temperature of the bullets to 200 K.

The ejection velocities of the bullets in CRL 618 are unknown, but their
lower limits can be estimated from the transverse velocities (proper
motions) of the optical lobes, $v_{\textrm{\scriptsize t}}$ \citep{R14} and
the radial velocities of the fast outflows, $v_{\textrm{\scriptsize r}}$
\citep{L13b}.  Table~\ref{tbl-1} shows the lower limits calculated with
$v=({v_{\textrm{\scriptsize t}}}^2+{v_{\textrm{\scriptsize r}}}^2)^{1/2}$. 
The ejection velocities $v_{\textrm{\scriptsize b}}$ must be higher than the
lower limits but not by much, because the bullets are much denser than the
surrounding.  Therefore, in our simulations, we set the ejection velocities
to be slightly higher.  As shown in Table \ref{tbl-2}, the ejection 
velocities $v_{\textrm{\scriptsize b}}$ are 330 and 350 km s$^{-1}$ for 
bullets 1 and 2, 280 km s$^{-1}$ for bullet 3, and 240 km s$^{-1}$ for 
bullets 4 and 5.

\begin{figure}
\centering{
\includegraphics[scale=0.45]{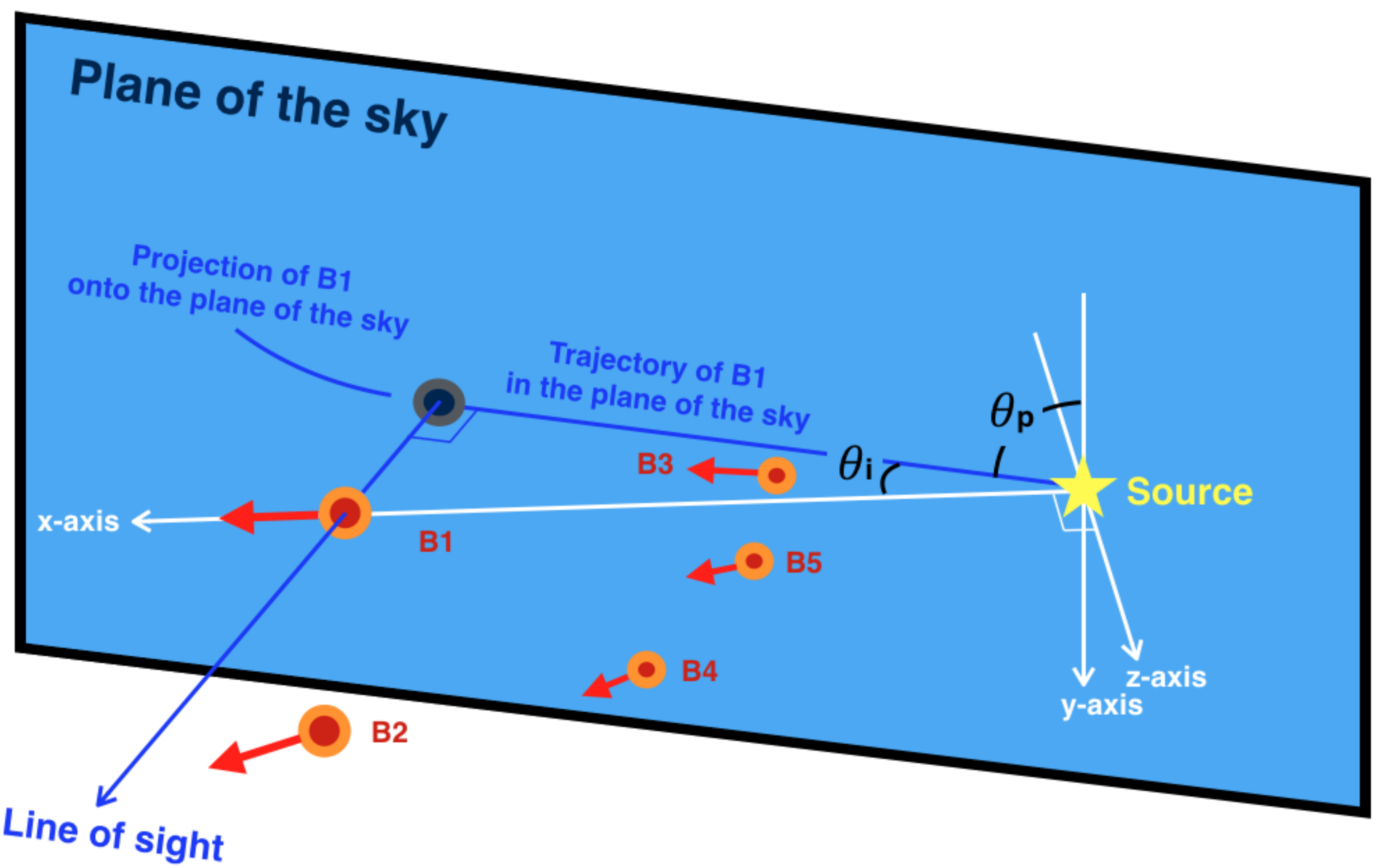}
}
\caption{Schematic diagram of five bullets ejected from the central source. 
The position angles $\theta_{\textrm{\scriptsize p}}$ of the bullets are set to 
those of the optical lobes and the molecular outflows of CRL 618. The 
inclination angles $\theta_{\textrm{\scriptsize i}}$ can be obtained from the 
proper motions of the optical lobes \citep{R14} and the radial velocities of the 
molecular outflows \citep{L13b}. 
Here we show the angles of bullet 1 (B1) as an example.  
\label{F2}}
\end{figure}

The ejection orientation of the bullets is defined by the position angle,
$\theta_{\textrm{\scriptsize p}}$, and the inclination angle,
$\theta_{\textrm{\scriptsize i}}=\textrm{tan}^{-1}(v_{\textrm{\scriptsize
r}}$/$v_{\textrm{\scriptsize t}})$, of the outflow axis of the optical lobes
and fast outflows, as listed in Table~\ref{tbl-1}.  For example, we assume bullet 1 to be ejected along the outflow axis of lobe E1 which has
$\theta_{\textrm{\scriptsize p}}=$ 90\degr\ and $\theta_{\textrm{\scriptsize
i}}=$ $-$24\degr\ (Fig. \ref{F2}).  This axis is also set to be the
$x$-axis in our simulations.  Since no fast outflow was observed at the tip
of lobe E2, the inclination angle of lobe E2 can not be determined.  Lobe E2
is assumed to have $\theta_{\textrm{\scriptsize i}}=$ $-$32\degr{}, lying
slightly in front of lobe E1.  The inclination angle of lobes E4 and E5 are
set to $-$35\degr\ and $-$45\degr{}, similar to the observed values.  For 
lobe E3, the transverse velocity and thus the inclination angle are 
uncertain.  We set $\theta_{\textrm{\scriptsize i}}=-$45\degr{} for this 
lobe, like lobe E5.

The bullet mass is set to $\sim$ 0.011 M$_{\sun}$ for bullets 1 and 2, and
0.004 M$_{\sun}$ for bullets 3, 4, and 5.  The total mass of the bullets is
therefore $\sim$ 0.034 M$_{\sun}$.  This value is about half of the total
mass (0.065 M$_{\sun}$) estimated from the total flux of the fast molecular
outflows in the east and west of CRL 618 \cite[with $|v|>20$ km
s$^{-1}$,][]{L13b}.  Therefore, this value is consistent with the
observations.  The density of the bullets is given by 
\begin{equation}
\rho_{\textrm{\scriptsize b}} =
\frac{M_{\textrm{\scriptsize b}}} { \pi ({d_{\textrm{\scriptsize b}}}/{2})^2 \, l_{\textrm{\scriptsize b}}}
\end{equation}
where $d_{\textrm{\scriptsize b}}$, $l_{\textrm{\scriptsize b}}$, and
$M_{\textrm{\scriptsize b}}$ are the diameter, length, and mass of the
bullets.  As mentioned earlier, $d_{\textrm{\scriptsize b}}$ and
$l_{\textrm{\scriptsize b}}$ are both 200 AU.  Therefore, the density of the
bullets is $\sim$ (0.4--1.0) $\times 10^{-15}$ g cm$^{-3}$, and thus the
number density of hydrogen nuclei in the bullets is
$\rho_\textrm{\scriptsize b}/(1.4m_{\textrm{\scriptsize H}}) \sim$
(1.6--4.4)$\times 10^{8}$ cm$^{-3}$ (Table~\ref{tbl-2}).  

In our model, the densities of the bullets are higher than those adopted in 
the previous simulations:  \citet{B13} adopted a number density of $4 \times 
10^4$ cm$^{-3}$ for the bullet in their model.  Unlike their model, there is a 
dense core present in our model.  \citet{R14} and \citet{V14} adopted an average 
density of $10^6$ cm$^{-3}$ for the jet in their precessing jet model.  In their 
model, a dense core is included, with a density of $\sim 10^6$ cm$^{-3}$ near 
the source.  In our model, the dense core has a density of $\sim 
10^8$ cm$^{-3}$ near the source (as described in the next section), which is 
$\sim$ 2 orders of magnitude higher than that in their precessing jet model.  
Thus, higher-density bullets are needed to excavate the holes in the dense core.  
As discussed earlier, the total mass of the bullets is 0.034 M$_{\sun}$, still 
consistent with the submillimeter observations.  

The mass-loss rate is given by 
\begin{equation}
\dot{M}_{\textrm{\scriptsize b}} = \pi (\frac{d_{\textrm{\scriptsize b}}}{2})^2 \, v_{\textrm{\scriptsize b}} \, \rho_{\textrm{\scriptsize b}}, 
\end{equation} 
which is $\sim 4 \times 10^{-3}$ M$_{\sun}$ yr$^{-1}$ for bullets 1 and 2, 
and $\sim 10^{-3}$ M$_{\sun}$ yr$^{-1}$ for bullets 3, 4, and 5.

{\subsection{The AGB Halo and the Dense Core}\label{CORE}}

The AGB halo is assumed to be spherical symmetric with a
constant mass-loss rate. Thus, the density of the AGB halo is given by 
\begin{equation}
\rho(r) = \frac{\dot{M}_{\textrm{\scriptsize a}}}{4 \pi r^2\, v_{\textrm{\scriptsize a}}}
\end{equation}
where $\dot{M}_{\textrm{\scriptsize a}}$ is the mass-loss rate assumed to be
$\sim$ 5.5$\times10^{-5}$ M$_{\sun}$ yr$^{-1}$ \cite[derived from the number 
density of molecular hydrogen
in Table 1 in][with Helium added]{S04}, $r$ is the radial distance to the 
central source, and $v_{\textrm{\scriptsize a}}$ is the
expansion velocity that was found to be $\sim$ 16 km s$^{-1}$ \citep{L13a}. 
Since the bullets move much faster than the AGB halo,
$v_{\textrm{\scriptsize a}}$ is set to zero for simplicity, after the
density of the AGB halo is calculated.  In observations, the halo of CRL
618 was found to have a temperature of 10--45 K \citep{S04}.  Here we set
the temperature of the AGB halo to be 10 K.  These two settings of the AGB
halo will not affect the dynamics of the outflow lobes to be produced.

\begin{figure}
\centering{
\includegraphics[]{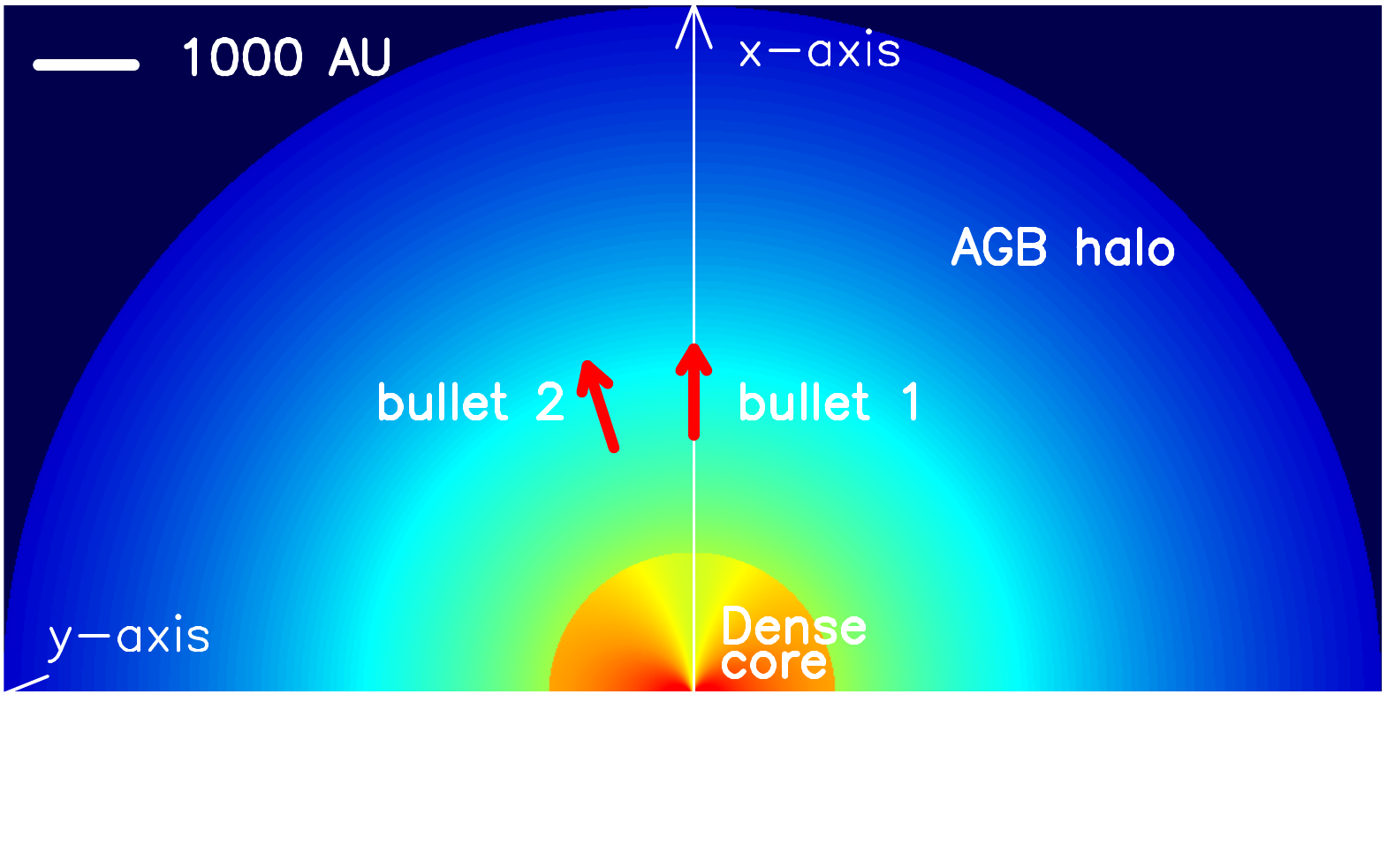}
}
\caption{Schematic diagram of the AGB halo and the toroidal dense core, with 
the first two bullets.  The dense core has a radius of 1500 AU.  
\label{F3}}
\end{figure}

The outer part of the dense core has been detected in CRL 618 with a radius
of $\sim$ 1100 AU \citep{S04}.  However, the actual radius of the dense core
should be larger.  Thus, in our simulations, the radius of the dense core
$R_{\textrm{\scriptsize c}}$ is assumed to be 1500 AU.  The density
distribution is assumed to be toroidal instead of spherical (Fig. 
\ref{F3}) because the dense core seems to be the inner part of the 
expanding torus detected in the equatorial plane extending to 5000 AU from 
the center \citep{SS04}.  For a toroidal density distribution, the simple 
form can be written as
\begin{equation}
\rho_{\textrm{\scriptsize c}}(r,\theta) = 
(A + B \, \mathrm{sin}^{2}\theta)\,\rho(r),\,\, \textrm{with}\,\,
\rho(r)=\frac{\dot{M}_{\textrm{\scriptsize c}}(r)}{4 \pi r^2\, v_{\textrm{\scriptsize c}}(r)}
\end{equation}
where $\theta$ is the angle measured from the polar axis, A is the isotropic
parameter, and B is the toroidal parameter.  The density structure would
become spherical when $A=1$ and $B=0$.  Here, for our toroidal 
structure, we assume $A\simeq 0.03$ and $B\simeq 2$, so that the density at 
$\theta=45$\degr\ is about half of that at the equator at the same distance.  
$\dot{M}_{\textrm{\scriptsize c}}(r)$ and $v_{\textrm{\scriptsize c}}(r)$
are the mass-loss rate and the expansion velocity of the dense core,
respectively.  The density profile in the $r$-axis, $\rho_{\textrm{\scriptsize c}}(r)$, is
unknown because existing observations do not have enough resolution to
resolve the dense core.  Here we assume $\rho_{\textrm{\scriptsize c}}(r)
\propto r^{-p}$.  If $p=2$ or 3, the density at the center
would become too high for the
bullets to penetrate, therefore we use $p=1$.  In the
observation of \citet{L13a}, $v_{\textrm{\scriptsize c}}(r)$ was found to be
linearly increasing with the radius,
\begin{equation}
v_{\textrm{\scriptsize c}}(r) = \left \{ \begin{array}{ll}
    v_a \left( \frac{r}{r_{\textrm{\scriptsize o}}} \right), \quad \mbox{if $r<r_{\textrm{\scriptsize o}}$} \\
    v_a, \quad \mbox{if $r_{\textrm{\scriptsize o}}<r<R_{\textrm{\scriptsize c}}$} 
    \end{array} \right.
\end{equation}
where $r_{\textrm{\scriptsize o}}=630$ AU.  After substituting
$v_{\textrm{\scriptsize c}}(r)$ into our density profile and rearranging, we
have
\begin{equation}
\dot{M}_{\textrm{\scriptsize c}} = \left \{ \begin{array}{ll}
    8\times10^{-4} \left( \frac{r}{r_{\textrm{\scriptsize o}}} \right)^2 \rm{M}_{\sun} \, \rm{yr}^{-1}, \quad \mbox{if $r<r_{\textrm{\scriptsize o}}$} \\
    2\times10^{-3} \left( \frac{r}{R_{\textrm{\scriptsize c}}} \right) \rm{M}_{\sun} \, \rm{yr}^{-1}, \quad \mbox{if $r_{\textrm{\scriptsize o}}<r<R_{\textrm{\scriptsize c}}$}. 
    \end{array} \right.
\end{equation}
Therefore, the mass-loss rate of the dense core is $\sim$ 40 times that of
the AGB halo at $r=R_{\textrm{\scriptsize c}}$, and it decreases to that of
the AGB halo at $r \simeq 170$ AU.  The mean value of the mass-loss rate of
the dense core is $\sim 10^{-3}$ M$_{\sun}$ yr$^{-1}$, similar to that
obtained by \citet{L13a}, which is $\sim 1.15 \times 10^{-3}$ M$_{\sun}$
yr$^{-1}$.

\section{RESULTS}

\subsection{Model 1: Two-Bullet Model}
 
\begin{figure}
\centering{
\includegraphics[scale=0.3]{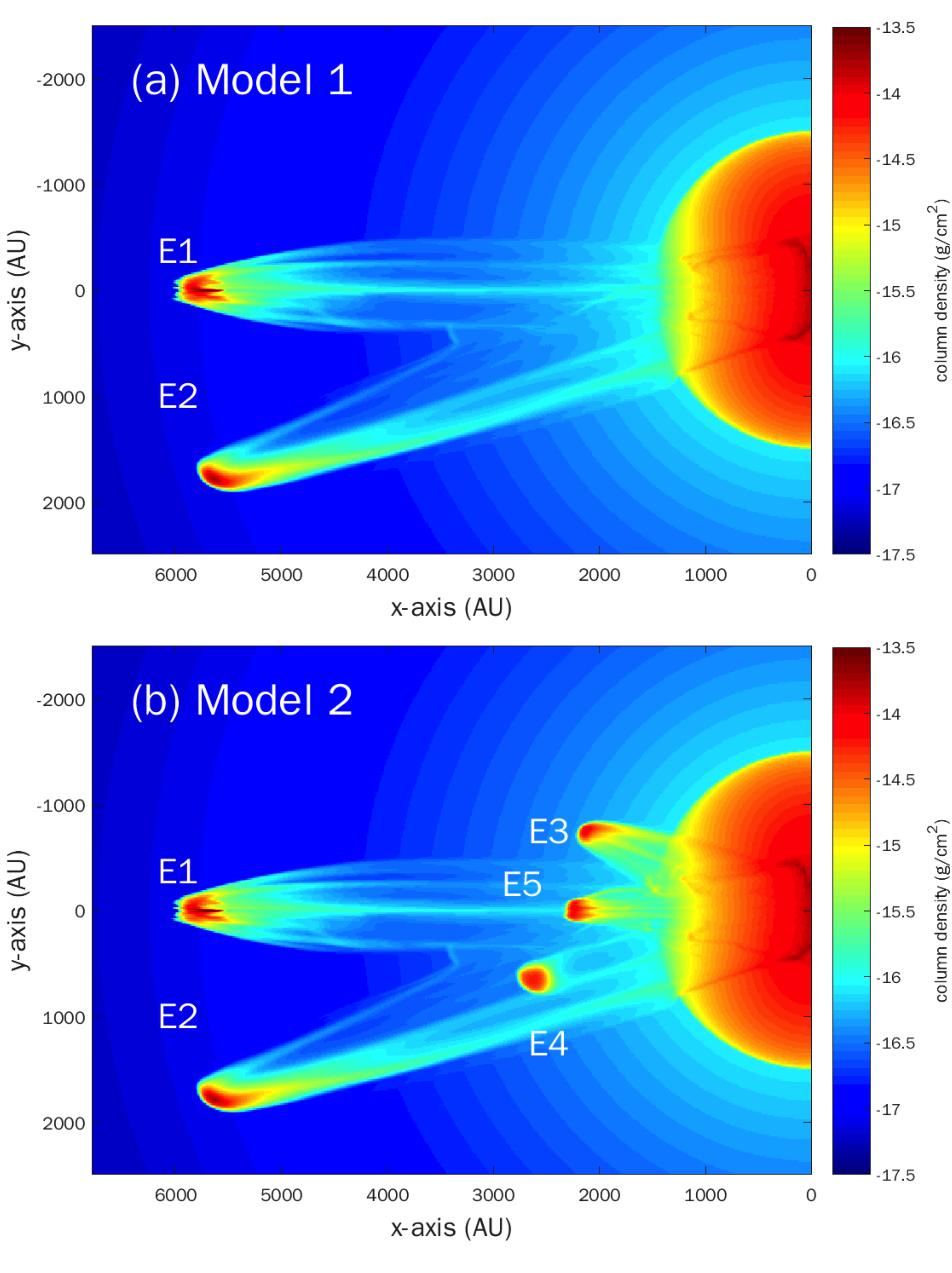}
}
\caption{Results of ZEUS-3D simulations of (a) Model 1 and (b) Model 2.  
The colors show the column density, integrated along the z-axis. 
Panel (a) shows the result of Model 1, including the E1 (position angle of 
90\degr , toward the east) and E2 (position angle of 108\degr , toward the 
southeast) lobes.  Panel (b) shows the result of Model 2, including the E1 
and E2 lobes, and also the shorter lobes E3 (toward the northeast), E4 
(toward the southeast), and E5 (toward the east). 
\label{F4}}
\end{figure}

Figure \ref{F4}a shows the column density in Model 1 at 90 yr.  As
the two bullets are ejected into the dense core and AGB halo, they produce
two collimated lobes, E1 and E2.  For these two lobes, the lengths are
$\sim$ 6000 AU and the transverse widths are $\sim$ 1000 AU, similar to the
observed lengths and widths of the two major lobes, E1 and E2, of CRL 618. 
The column density shows that the majority of the bullet mass is still at
the tips, which could be traced by the fast molecular outflows \citep{L13b}.

\subsection{Model 2: Five-Bullet Model}

In Model 2, five bullets are ejected in two episodes, producing the five
eastern lobes of CRL 618 (Fig. \ref{F4}b).  In the first episode, two
bullets are ejected, producing lobes E1 and E2.  In the second episode
($\sim$ 40 yr after the first episode), three bullets are ejected, producing
lobes E3, E4, and E5.  Lobe E3 has an inclination angle of $-$45\degr\ in
the northeast, and it is the smallest lobe with a length of $\sim$ 2200
AU, as seen in the observation.  The lengths of lobes E4 and E5 are $\sim$ 
2800 AU and 2300 AU, respectively. A U-shaped cavity wall is seen at the 
base, consisting of swept-up material in the dense core.

\subsection{Comparison with CO Observations}

Since Model 2 can reproduce the structures of the five eastern lobes of CRL
618, here we choose this model to be compared with the observations.  Since
CO $J=3$--2 has been observed at very high resolution of $\sim$
\arcsa{0}{5} and found to trace the fast molecular outflows reasonably
well \citep{L13b}, we calculate the CO $J=3$--2 intensity maps and the
position-velocity (PV) diagrams from Model 2 and compare them to the
observations of the molecular outflows in the east of CRL 618.  We use a
radiative transfer code with an assumption of local thermal equilibrium
(LTE) to calculate the CO emission.  The x-axis is tilted toward us from the
plane of the sky by 24\degr{}.  The systemic velocity is $-$21.5 km
s$^{-1}$, as adopted for CRL 618 in \citet{L13a}.

\begin{figure}
\centering{
\includegraphics[scale=0.4]{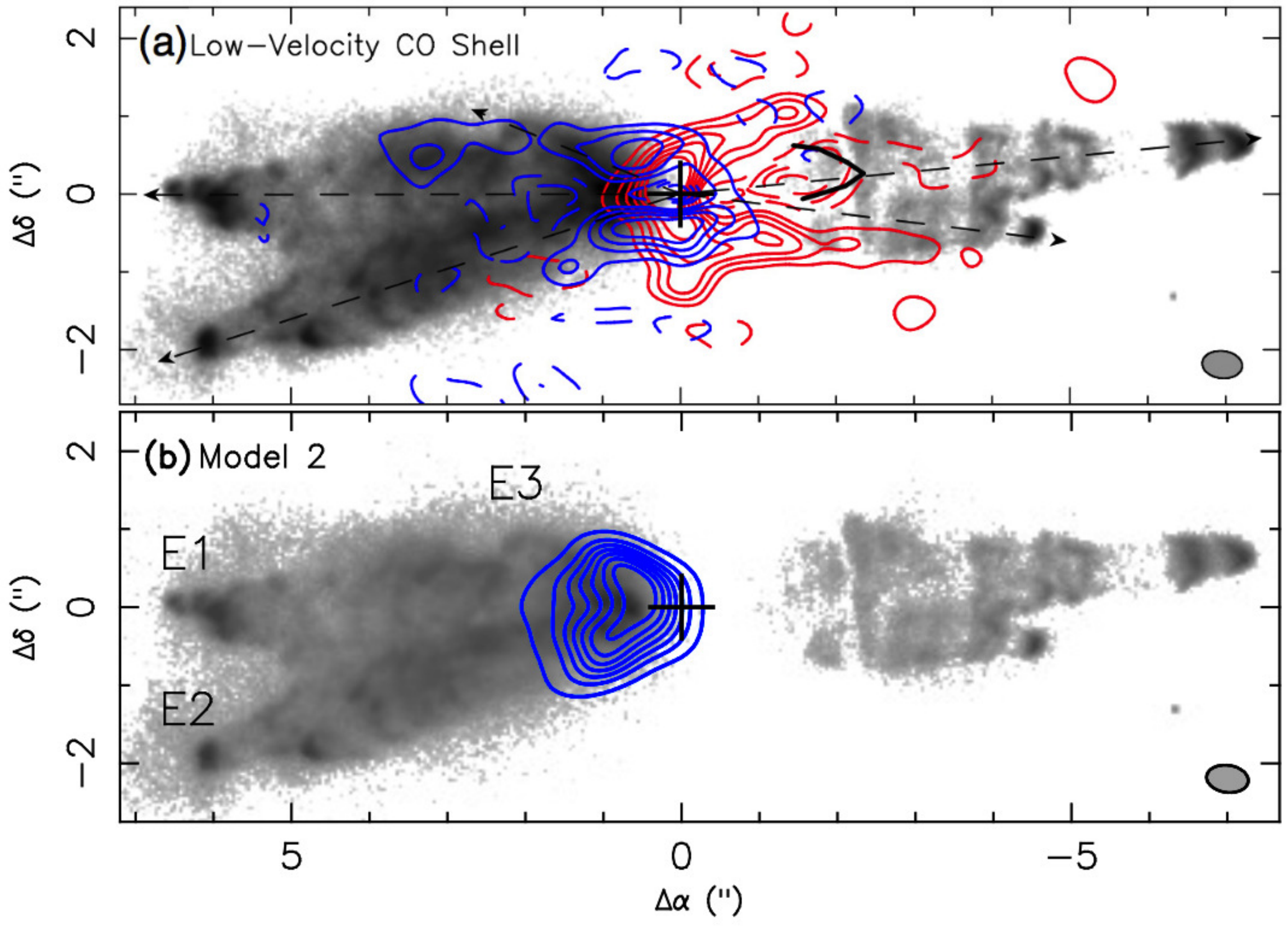}
}
\caption{
Panel (a) shows the observed low-velocity (LV) CO $J=3$--2
emission (contours) plotted on the H$\alpha$ image of CRL 618 (Figure 2 of
\citet{L13b}).  The LV CO contours show the U-shaped cavity walls in the
redshifted (averaged from $-$16.9 to $-$8.5 km s$^{-1}$) and blueshifted
(averaged from $-$35.2 to $-$26.8 km s$^{-1}$) emission.  The systemic
velocity is $-$21.5 km s$^{-1}$.  The contour levels start at $\sim$ 50 mJy
beam$^{-1}$ with a step of $\sim$ 100 mJy beam$^{-1}$.  The synthesized beam
is $\sim$ \arcsa{0}{53} $\times$ \arcsa{0}{35} at a position angle of $\sim$
83\degr .  Panel (b) shows the simulated CO $J=3$--2 emission in the
low velocity (averaged from $-$26 to $-$22 km s$^{-1}$), showing the cavity 
wall in Model 2.  The contour levels and the synthesized beam are the same as 
in Panel (a).  
\label{F5}}
\end{figure}

Figure \ref{F5} shows the comparison of low-velocity (LV) CO $J=3$--2
emission between the observation and the model for the east side of CRL 618. 
The figures are plotted over the $HST$ H$\alpha$ image of CRL 618
\citep{L13b} in order to show the relative distribution of the molecular
outflows along the optical lobes.  In the observation (Fig. \ref{F5}a), 
the low-velocity emission shows a U-shaped cavity wall encompassing the two
major lobes, with the northern part around the northern edge of the E1 lobe
and the southern part around the southern edge of the E2 lobe, extending
away from the central source.  A similar cavity wall is seen around the
western lobe as well.  These structures are also seen in the lower transition 
line of CO at $J=2$--1 \citep{S04}.

In the model (Fig. \ref{F5}b), the low-velocity emission also shows a
U-shaped cavity wall encompassing the E1 and E2 lobes,  
extending to $\sim$ \arcsa{1}{7} from the source.  The emission
is from the swept-up material in the dense core, which has a radius of 1500
AU ($\sim$ \arcsa{1}{7}).  Unlike the observation, however, no emission is
seen extending over $\sim$ \arcsa{1}{7} from the source to $\sim$ \arcs{4}
to the northeast.  In our model, that region is within the AGB halo.  Thus, 
in order to produce detectable emission there, the AGB halo could be denser 
than currently assumed in our model.

\begin{figure}
\centering{
\includegraphics[scale=0.4]{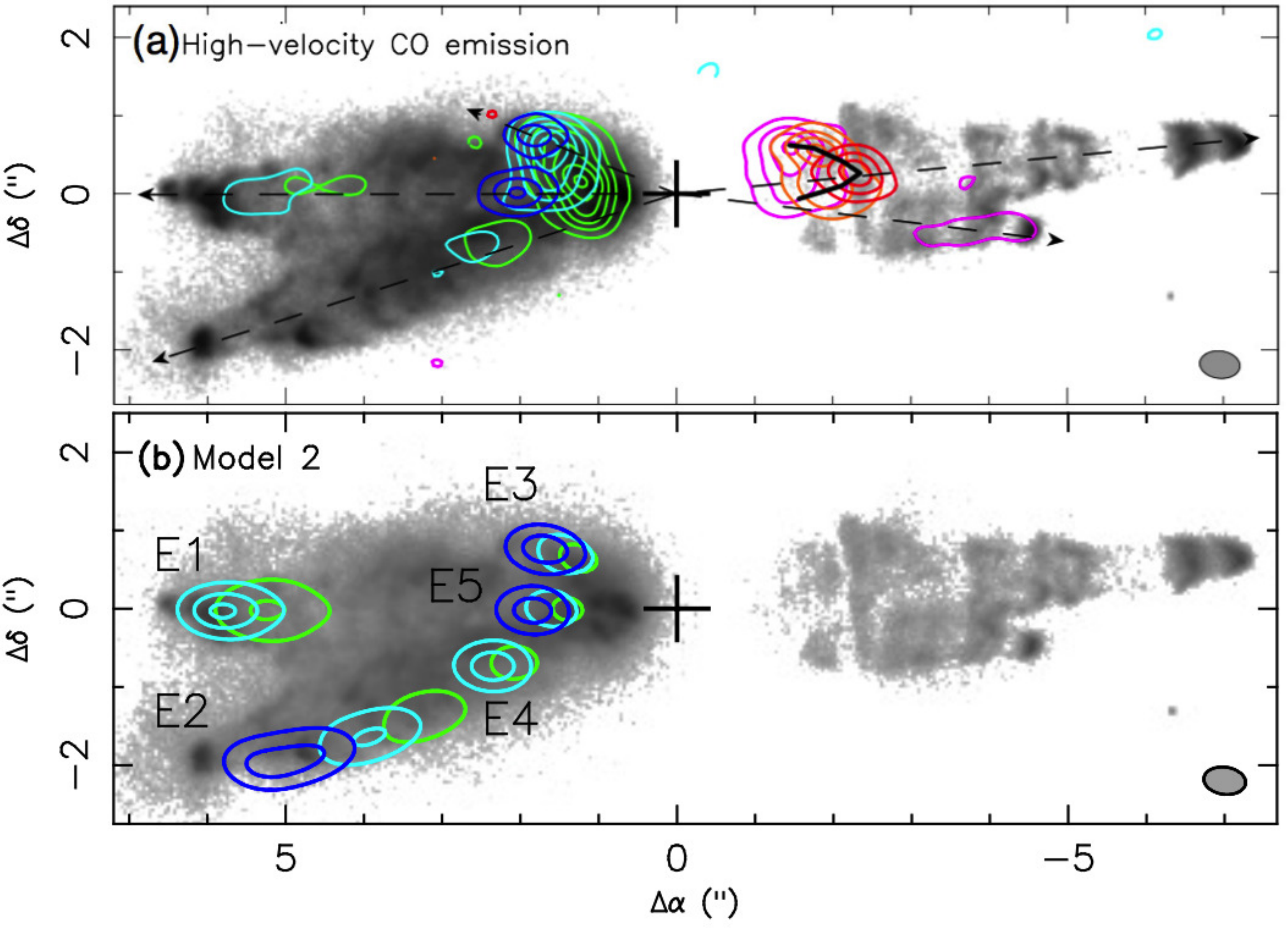}
}
\caption{Panel (a) shows the observed high-velocity (HV) CO $J=3$--2
emission (contours) plotted on the H$\alpha$ image of CRL 618 (from Figure 2
in \citet{L13b}).  The velocity ranges of the contours are: (blue) $-$183.2
to $-$160.6, (cyan) $-$157.8 to $-$132.5, and (green) $-$132.5 to $-$114.1
km s$^{-1}$.  The contours start at 1 Jy beam$^{-1}$ km s$^{-1}$ with a step
of 2 Jy beam$^{-1}$ km s$^{-1}$.  The synthesized beam is $\sim$
\arcsa{0}{53} $\times$ \arcsa{0}{35} at a position angle of $\sim$
83\degr{}.  Panel (b) shows the simulated HV CO $J=3$--2
emissions (contours) in Model 2, with similar velocity ranges to those in
Panel (a).  HV CO emissions are seen at the tips of the E1 to E5
lobes.  The contours start at 0.5 Jy beam$^{-1}$ km s$^{-1}$ with a step of
1 Jy beam$^{-1}$ km s$^{-1}$.  The synthesized beam is the same as that in 
Panel (a).
\label{F6}}
\end{figure}

Figure \ref{F6} shows the comparison of high-velocity (HV) CO $J=3$--2
emissions between the observation and the model.  Both panels are also
plotted over the $HST$ images of CRL 618.  In the observation (Fig.
\ref{F6}a), high-velocity ($-$183.2 to $-$114.1 km s$^{-1}$) emission is
observed at the tips of lobes E1, E3, E4, and E5, but not observed at the tip
of lobe E2.  In the model, high-velocity emission, with a similar velocity
range to that in the observation, is predicted to be seen at the tips
of all the lobes.  The velocity of the emission increases with distance,
similar to that seen in the observation.  Notice that the emission intensity
in our model is about half of that observed and thus more bullet mass may be
needed to produce more HV CO emission.  The reason for no HV emission being
detected at the tip of and along the southern edge of lobe E2 is unknown.
 
\begin{figure}
\centering{
\includegraphics[scale=0.6]{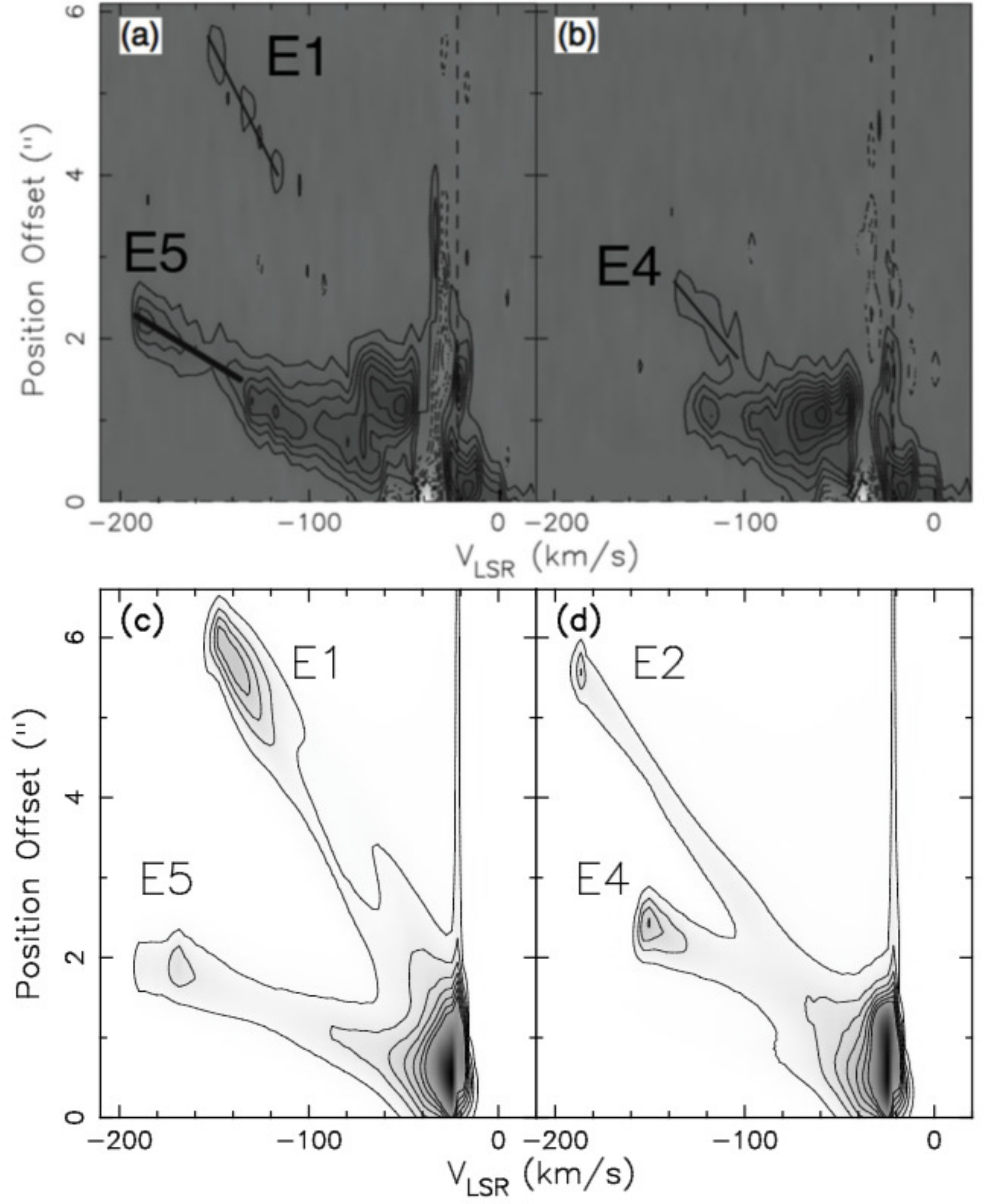}
}
\caption{
Panel (a) and (b) show the position-velocity (PV) diagrams of CO $J=3$--2 
emission of CRL 618, along the E1 and E2 lobes, respectively (from Figure 5 
of \citet{L13b}).  The velocities linearly increase with the distance to the 
central source.  The contours start from 40 Jy beam$^{-1}$ with a step of 80 
Jy beam$^{-1}$.  Panel (c) and (d) shows the simulated CO emission produced 
by bullets 1 and 5, and bullets 2 and 4, respectively.  The contours start at 
20 Jy beam$^{-1}$ with a step of 40 Jy beam$^{-1}$. 
\label{F7}}
\end{figure}

Figures \ref{F7}a and \ref{F7}b show the PV diagrams of the observed
outflows in CO, cut along the axis of lobes E1 and E2.  Figures
\ref{F7}c and \ref{F7}d show the same PV diagrams but calculated from the
model.  In these diagrams, lobes E4 and E5 can also been seen.  In  
Figures \ref{F7}c and \ref{F7}d, the bullets are seen at the tips of the 
outflows and thus are traced by the fast molecular outflows.  The diagrams 
also show that the velocities of the molecular outflows decrease linearly 
with the distance to the central source, as seen in the observations.  The 
slope of the PV structure is roughly equal to $p/v_{\textrm{\scriptsize
r}}=t/{\mathrm{tan}\theta_{\textrm{\scriptsize i}}}$, where $p$ is the
projected distance from the center, and $t$ is the time after ejection. 
Therefore, different slopes are seen for the outflows in different ejection
episodes.  Notice that since no molecular emission is detected for lobe
E2, no PV structure is detected for lobe E2 in the observation, as
shown in Figure \ref{F7}b.

\section{DISCUSSIONS}

\subsection{Physical Properties of the Dense Core}


In order to test our bullet models realistically, the physical parameters of
the dense core are set to be close to those derived from the observations. 
The dense core is assumed to be toroidal, because the dense core seems to be
the inner part of the expanding torus detected in the equatorial plane
extending to 5000 AU from the center \citep{SS04}.  In addition, bipolar and
multipolar PPNe are often seen with dense tori at their waists
\citep{K98,S98,V07,S08}.  The dense core is assumed to have a mean mass-loss
rate of $\sim$ $10^{-3}$ M$_{\sun}$ yr$^{-1}$, similar to the observed
values [$\sim$ 0.7 $\times$ $ 10^{-3}$ M$_{\sun}$ yr$^{-1}$ in the outer
part \cite[][with Helium added]{S04} and 1.2$\times$ $10^{-3}$ M$_{\sun}$ 
yr$^{-1}$ in the inner part \citep{L13a}].

On the other hand, the density profile of the dense core is uncertain and
assumed to be proportional to $r^{-1}$ in our models for the first attempt.  
In the observations, the outer part of the dense core was observed with an 
angular resolution of $\sim$ \arcsa{1}{1} and thus was not resolved 
\citep{S04}.  The density profile there was assumed to be proportional to 
$r^{-2}$ to model the observed emission.  The inner part of the dense core 
was observed with an angular resolution of $\sim$ \arcsa{0}{5} and thus was 
also not resolved \citep{L13a}.  The density profile there was assumed to be
proportional to $r^{-3}$, in order to have a constant mass-loss rate.  These
two density profiles are steeper than the density profile used in our
models.  If we used those steeper density profiles, the density in the
innermost part of the dense core would be higher and thus more massive
bullets would be needed to penetrate the dense core.  Observations at higher
angular resolution with the Atacama Large Millimeter/submillimeter Array
(ALMA) are needed to resolve the dense core in order to better constrain the
density profile of the dense core and thus the parameters of the bullets.

\subsection{Multipolar Features: Precession or Bullet Ejections?}

Precessing jet models have been used to reproduce the multipolar structures 
of CRL 618.  However, a standard precessing jet model would produce
point-symmetric multipolar lobes about the central source \citep{R14},
inconsistent with the observations, which show asymmetric multipolar lobes
about the central source.  In order to produce the asymmetric multipolar
lobes, \citet{V14} adopted an asymmetrical jet-ejection mechanism for the
precessing jet model.  In this case, the morphology and the kinematics of
the lobes can be in reasonable agreement with the observations. 
Nonetheless, the jet-ejection mechanism is complicated, requiring an
alternation in the ejections of the jet from two sides of the precessing
accretion disk around a binary companion.

Here, we extend the one-bullet model studied earlier in \citet{D08} and
\citet{B13} to the multi-directional bullet model in order to produce the
multipolar lobes in CRL 618.  In order to have a more realistic comparison
with the observations, we also include the dense core detected recently in
CRL 618 \citep{L13a} and set the physical parameters of our bullets using
the fast molecular outflows recently detected in \citet{L13b}.  As discussed
earlier, our model can reproduce the morphology and the kinematics of the
multipolar lobes in the east of CRL 618 reasonably well.  Since the lobes in
the west are different from those in the east, a different set of bullets are
needed to produce them.  Notice that two episodes of bullet ejections are
assumed in our model, based on the observations of fast molecular outflows
in \citet{L13b}.  However, since all the optical lobes were found to have
similar dynamical ages \citep{B13}, the bullets could also be ejected at the
same time.  Further observations are needed to determine this.

Although both of the precessing jet model and our multi-directional bullet 
model can produce the multiple lobes, there is an observational difference for 
the shock structures at the tips of the lobes between the two models.  As seen 
in Figure 4 of \citet{V14}, the synthetic image shows that the shapes of the 
shocks produced by precessing jets are blunt (U-shaped).  In our bullet model, on 
the other hand, the shapes at the tips of the lobes are sharp (V-shaped).  In the 
observations of CRL 618, the shapes of the shocks are as sharp as those seen in 
our simulations.

\subsection{The Origin of Bullets}



The origin of the bullets is still uncertain.  \citet{D08} argued that the
bullets may be driven by an explosive magneto-hydrodynamic (MHD) mechanism
\citep[see][]{M06}.  Near the end of the AGB phase, the central star rotates
very fast so that the magnetic field can be highly twisted near the central
star.  If the magnetic field is strong enough, the magnetic pressure force
can drive a magneto-rotational explosion.  Such an explosion event could
produce multiple bullets simultaneously, producing the multipolar lobes of
CRL 618.  Further study is needed to determine if this kind of explosion can
occur twice, as needed in our simulations.

Interestingly, extremely low isotope ratios of $^{12}$C/$^{13}$C ($\sim$ 10)
and $^{14}$N/$^{15}$N (tentatively found to be $\sim$ 150) were found in the
dense core at the center \citep{L13a}.  These low isotope ratios are
unexpected to be seen in a C-rich dense core \citep{M09,Palmerini2011}, such
as that in CRL 618.  As suggested in \citet{L13a}, these low ratios could be
due to a hot CNO cycle as in a nova-like explosion.  If this is the case,
then the bullets could be ejected by these explosions.  Theoretically, a
nova requires a close binary system with the secondary being a white dwarf,
and it is triggered by the accretion of hydrogen gas from the primary onto
the white dwarf.  The hydrogen gas can be pulled onto the surface of the
white dwarf and then form an envelope massive enough to ignite the hot CNO
cycles.  If this is the case in CRL 618, then the center would be a binary
system.  This binary system in turn could produce the flattened common
envelope around the center, seen as the expanding torus.

In our model, there are two episodes of bullet ejections. The first episode
occurred about 90 yr ago, and the second episode occurred about 50 yr ago. 
Interestingly, by studying the HII region at the center, \citet{T13} found
that CRL 618 turned from a PPN into a PN at the center about 50 yr ago. 
Thus, it is possible that the second episode of bullet ejection is related
to the formation of the HII region at the center and thus the evolutionary
phase change of the central star.


\section{CONCLUSIONS}

With hydrodynamical simulations, we have studied the shaping mechanism of
the multipolar PPN CRL 618 using a multi-directional bullet model.  In our
simulations, we include a spherical AGB halo with a mass-loss rate of $\sim$
$5.5\times10^{-5}$ M$_{\sun}$ yr$^{-1}$ and a toroidal dense core with a
mean mass-loss rate of $\sim$ $10^{-3}$ M$_{\sun}$ yr$^{-1}$, as found in
previous observations.  Also, based on previous observations of fast
molecular outflows in the east of the central source, we assume two episodes
of bullet ejection, with the first at $\sim$ 90 yr ago and the second at
$\sim$ 50 yr ago, and with a total mass of $\sim$ 0.034 M$_{\sun}$.  

We find that our simulations can reproduce the morphology of the multipolar
lobes in the east of CRL 618.  The simulated CO emissions show the structure
and the kinematics of the molecular outflows similar to those seen in the
observations.  At low velocity, the emission shows a similar cavity wall to
that seen in the observations, surrounding the two major lobes in the east
extending to $\sim$ \arcsa{1}{7} from the source.  However, in order to
produce the observed emission extending from $\sim$ \arcsa{1}{7} to \arcs{4},
the AGB halo might need to be denser than that assumed here.  At high 
velocity, the emission is distributed at the tips of all the lobes.  The 
velocity increases with the distance, similar to that seen in the 
observation.  Since the intensity is about half of that observed, more massive 
bullets might be needed. 

The origin of the bullets is unknown. The ejection of these bullets could be
due to magneto-rotational explosions or nova-like explosions around a binary
companion.  Interestingly, the more recent ejection at about 50 yr ago could
be related to the formation of the HII region at the center when CRL 618
turned from a PPN into a PN.

\acknowledgements

Use of ZEUS-3D, developed by D. A. Clarke at the Institute for Computational 
Astrophysics (http://www.ica.smu.ca) with financial support from the Natural 
Sciences and Engineering Research Council of Canada (NSERC), is hereby 
acknowledged.
P.-S.H, C.-F.L, and A. M acknowledge grants from the National Science
Council of Taiwan (NSC 101-2119-M-001-002-MY3 and MoST
104-2119-M-001-015-MY3) and the Academia Sinica (Career Development Award). 
The authors acknowledge the support of the Theoretical Institute for Advanced 
Research in Astrophysics (TIARA) in Academia Sinica Institute of Astronomy 
and Astrophysics (ASIAA).  P.-S.H. is grateful to Bruce Balick, Mark Morris, 
and Raghvendra Sahai for the valuable discussions during the PRCSA meeting.

\appendix

\clearpage

\newpage

\clearpage

\begin{table}
\caption{Observation values of the optical lobes 
\citep[$^{\textrm{\scriptsize a}}$][]{R14} and CO outflows of CRL 618 
\citep[$^{\textrm{\scriptsize b}}$][]{L13b}.\label{tbl-1}}
\begin{tabular}{cccccccc}

\tableline\tableline
Lobe & 
Projected length & 
$v_{\textrm{\scriptsize t}} \, ^{\textrm{\scriptsize a}}$ & 
$v_{\textrm{\scriptsize r}} \, ^{\textrm{\scriptsize b}}$ & 
$v=({v_{\textrm{\scriptsize t}}}^2+{v_{\textrm{\scriptsize r}}}^2)^{1/2}$ & 
${\theta}_{\textrm{\scriptsize p}}$ & 
${\theta}_{\textrm{\scriptsize i}}$ & 
Kinematic age \\

 & & (km s$^{-1}$) & (km s$^{-1}$) & (km s$^{-1}$) & & & (yr) \\

\tableline
E1 & \arcsa{6}{8} & $288 \pm 18$ & 125 & $314 \pm 17$ & 90\degr & $-$24\degr &  100 \\
E2 & \arcsa{6}{3} & $243.3 \pm 14$ & \nodata & \nodata & 108\degr & \nodata & 110 \\
E3 & \arcsa{2}{3} & $84 \pm 18$ & 140 & $163 \pm 10$ & 70\degr & $-$59\degr & 116 \\
E4 & \arcsa{2}{5} & $158 \pm 18$ & 110 & $193 \pm 15$ & 104\degr & $-$35\degr &  68 \\
E5 & \arcsa{2}{3} & $158 \pm 18$ & 165 & $228 \pm 13$ & 90\degr & $-$46\degr &  62 \\

\tableline\tableline
\end{tabular}
\end{table}

\begin{table}
\caption{Bullet parameters in our models.\label{tbl-2}}
\begin{tabular}{cccccccc}
\tableline\tableline

 & 
$M_{\textrm{\scriptsize b}}$ & 
$d_{\textrm{\scriptsize b}}$ & 
$l_{\textrm{\scriptsize b}}$ & 
$v_{\textrm{\scriptsize b}}$ & 
$n_{\textrm{\scriptsize H}}$ & 
${\theta}_{\textrm{\scriptsize p}}$ & 
${\theta}_{\textrm{\scriptsize i}}$ \\

 & (M$_{\sun}$) & (AU) & (AU) & (km s$^{-1}$) & (cm$^{-3}$) \\
   
   \tableline

   Bullet 1 & 0.011 & 200 & 200 & 330 & 4.4 $\times 10^{8}$ & 90\degr & $-$24\degr \\
   Bullet 2 & 0.011 & 200 & 200 & 350 & 4.4 $\times 10^{8}$ & 108\degr & $-$32\degr \\
   Bullet 3 & 0.004 & 200 & 200 & 280 & 1.6 $\times 10^{8}$ & 70\degr & $-$45\degr \\
   Bullet 4 & 0.004 & 200 & 200 & 240 & 1.6 $\times 10^{8}$ & 104\degr & $-$35\degr \\
   Bullet 5 & 0.004 & 200 & 200 & 240 & 1.6 $\times 10^{8}$ & 90\degr & $-$45\degr \\

\tableline\tableline
\end{tabular}
\end{table}

\end{document}